%% file: main.tex
\documentclass[conference]{IEEEtran}

\newcommand{\EBY}[1]{{\color{black}#1}}

\usepackage{pack}

\allowdisplaybreaks

\begin{document}

	\title{List Message Passing Decoding of Non-binary
		Low-Density Parity-Check Codes}
\author{\IEEEauthorblockN{Emna Ben Yacoub}
	\IEEEauthorblockA{Institute for Communications Engineering \\
		Technical University of Munich, Germany\\
		Email: emna.ben-yacoub@tum.de}}

	\markboth{}{}%

	
	\maketitle
		\begin{abstract}
		A decoding algorithm for $q$-ary low-density parity-check codes over the $q$-ary symmetric channel is introduced. The exchanged messages are lists  of symbols from $\Fq$.
		A density evolution analysis for maximum  list sizes $1$ and $2$ is developed. Thresholds for selected regular low-density parity-check code ensembles are computed showing gains with respect to \EBY{a similar} algorithm  in the literature. Finite-length simulation results \EBY{confirm}	the asymptotic analysis. 
		
	\end{abstract}
	
\input{./tex_files/acron}
\input{./tex_files/intro}

\input{./tex_files/prelim}

\input{./tex_files/lmp_description}
\input{./tex_files/lmp_DE_1}
\input{./tex_files/lmp_DE_2}
\input{./tex_files/results}


\input{./tex_files/conclusions}


\begin{appendices}
\end{appendices}

\clearpage
\bibliographystyle{IEEEtran}
\bibliography{IEEEabrv,confs-jrnls,references}



\end{document}

%% file: tex_files/acron.tex
\begin{acronym}
\acro{BEC}{binary erasure channel}
\acro{BSC}{binary symmetric channel}
\acro{BMP}{binary message passing}
\acro{BP}{belief propagation}
\acro{DE}{density evolution}
\acro{LDPC}{low-density parity-check}
\acro{MDPC}{moderate-density parity-check}
\acro{SC-LDPC}{spatially-coupled low-density parity-check}
\acro{lv}{$L$-value}
\acro{llv}{$\bm L$-vector}
\acro{LLR}{low-likelihood ratio}
\acro{QSC}{$q$-ary symmetric channel}
\acro{QEEC}{$q$-ary error and erasure channel}
\acro{SER}{symbol error rate}
\acro{SMP}{symbol message passing}
\acro{VN}{variable node}
\acro{CN}{check node}
\acro{RV}{random variable}
\acro{SC-LDPC}{spatially coupled low-density parity-check}
\end{acronym}

%% file: tex_files/intro.tex
\section{Introduction}\label{sec:Intro}
Non-binary \ac{LDPC} codes have shown an outstanding \EBY{error correction capability,} outperforming their binary counterparts substantially at short block lengths \cite{Davey_nonbin}. Nevertheless, the complexity of the \ac{BP} decoder for these codes  is  high. Various works  reduced the decoding complexity of non-binary \ac{LDPC} codes over \EBY{the} \ac{biAWGN} channel \cite{barnault_fast_2003,declercq_decoding_2007,Zhang_nonbin,Zha_EMS,wany_EMS,declerqNB} and \EBY{the} \ac{QSC} \cite{Luby_verif,matuz_verif,Shokrollahi_qsc,pfister_qsc,smp_lazaro,kurkoski_list}. In  \cite{declercq_decoding_2007}, an extension of the  min-sum algorithm to non-binary fields was presented.
The authors of \cite{Luby_verif} introduced a verification-based decoding algorithm for \ac{LDPC} codes over large alphabet on the \ac{QSC}. Similar approaches were considered in \cite{Shokrollahi_qsc}, \cite{PfisterLMP}.
It has been shown in \cite{Wang_verif} that \ac{LDPC} codes on the \ac{QSC} for large $q$ can approach the Shannon limit. In \cite{kurkoski_list}, the authors introduced a decoding algorithm over the \ac{QSC} (for any $q$) where the exchanged messages are  sets of symbols from $\Fq$ (including the empty set).
The performance of this algorithm improves with the list size, but this comes at the cost of an increasing data flow \EBY{within the decoder, with respect to the case where check and variable nodes exchange only hard decisions (i.e., symbols in $\Fq$)}.

 Inspired by \cite{kurkoski_list}, we propose   a message passing algorithm for $q$-ary \ac{LDPC} codes  over the \ac{QSC}, referred to as \ac{SRLMP}, where the variable and check nodes exchange a set of symbols from $\Fq$ of size at most $\Gamma$. The difference between our algorithm and the one in \cite{kurkoski_list} is the \ac{VN} update rule. In \cite{kurkoski_list}, the channel and the \ac{CN} messages \EBY{are mapped to binary vectors of length $q$. The entry corresponding to the symbol $a \in \Fq$ is one if the message contains the symbol $a$ and it is zero otherwise. These vectors} are summed at the \ac{VN} decoder. Whereas, for \ac{SRLMP} the channel and \ac{CN} messages are converted to log-likelihood vectors at the \acp{VN} by modeling the  extrinsic channel as a \ac{DMC}  whose transition probabilities may be estimated via \ac{DE}. This technique is similar to the \ac{TMP} decoder \cite{benyacoub_tmp} for binary codes, \EBY{and lays its foundation in the \ac{BMP} algorithm originally proposed in \cite{lechner_analysis_2012}}. 

In this work, we describe the \ac{SRLMP} decoder and develop the exact \ac{DE} for $\Gamma\in \{1,2\}$. We provide the iterative decoding thresholds of 
some regular \ac{LDPC} ensembles.

%% file: tex_files/prelim.tex
\section{Preliminaries}\label{sec:Prelim}

\subsection{Q-ary Symmetric Channel}

Consider a \ac{QSC} with error probability $\proberror$, input alphabet $\inputalphabet$ and output alphabet $\outputalphabet$, with  ${\inputalphabet = \outputalphabet = \{0, \alpha^0, \hdots, \alpha^{q-2}\}}$, where $\alpha$ is a primitive element of $\Fq$. 
Denote by $x \in \inputalphabet$  and $y \in \outputalphabet$ the channel input and channel output, respectively.
The transition probabilities of a \ac{QSC} with error probability $\proberror$ are
\begin{equation}\label{eq:QSC_trans_prob}
P(y|x) =  \begin{cases}
1 - \proberror & \mbox{if } \outputrvreal = \inputrvreal \\
\proberror / (q-1) & \mbox{otherwise}.
\end{cases} 
\end{equation}
The capacity of the \ac{QSC}, in symbols per channel use, is
\begin{equation}\label{eq:capacity_q}
C= 1 + \proberror \log_q \frac{\proberror}{q-1} + (1-\proberror) \log_q (1-\proberror).
\end{equation}

\subsection{Log-Likelihood Vector}
For a given channel output $y$ of a \ac{DMC} with input alphabet $\cX=\Fq$,  we introduce the normalized log-likelihood vector, also referred to as \acl{llv},
	\begin{equation}\label{eq:llvector}
	\bm{L}(y)=\left[ L_0(y), L_1(y), \ldots , L_{\alpha^{q-2}}(y)\right]
	\end{equation}
	whose elements are defined as
	\begin{align}\label{eq:llvector2}
	L_\genfq(y)=& \log \left(P(y|\genfq) \right) \quad \forall u \in \Fq.
	\end{align}
	The \acl{llv}  will be fundamental to the development of a  decoding algorithm for non-binary \ac{LDPC} codes over the \ac{QSC}. In particular, we will focus on  a message passing decoding algorithm where the  exchanged messages are list of symbols  from $\Fq$ of size at most $\Gamma$. In this case, a message sent from a \ac{CN} to a \ac{VN} can be modeled as the observation of the \ac{RV} $X$ after transmission over a $q$-ary input $|\cM_{\Gamma}|$-ary output discrete memoryless extrinsic channel \cite[Fig.~3]{ashikhmin_EXIT}, where $\cM_{\Gamma}$ is the message alphabet. In the decoding algorithm, we will use the 
	\acl{llv} $\bm{L}(y)$ of the communication channel observation and the \aclp{llv} of the \ac{CN} messages $\bm{L}(m), \forall m \in \cM_\Gamma$. While the transition probabilities of the communication channel, which is a \ac{QSC} with error probability $\epsilon$, are given in \eqref{eq:QSC_trans_prob},  the transition probabilities of the extrinsic channel are in general unknown but accurate estimates can be obtained via \ac{DE} analysis, as suggested in \cite{lechner_analysis_2012}.

\subsection{Non-binary LDPC Codes}
Consider next non-binary \ac{LDPC} codes defined by an $m \times n$ sparse parity-check matrix $\bm{H}$ with coefficients in $\Fq$. The parity-check matrix can be represented by a Tanner graph with $n$ \acp{VN} corresponding to codeword symbols and $m$ \acp{CN} corresponding to parity checks.  The edge label associated to the edge connecting  $\vn$ and  $\cn$ is denoted by $h_{\vn,\cn}$, with $h_{\vn,\cn} \in \Fqz$. The sets $\cN(\vn)$ and $\cN(\cn)$ denote the neighbors of \ac{VN} $\vn$ and \ac{CN} $\cn$, respectively. The degree  of a \ac{VN} $\vn$  is the cardinality of the set $\cN(\vn)$. Similarly, the degree of a \ac{CN} $\cn$  is the cardinality of the set $\cN(\cn)$.
The \ac{VN} edge-oriented degree distribution polynomial of an \ac{LDPC} code graph is given by 
$
    \lambda(x)= \sum_{i}\lambda_{i}x^{i-1}
$
where $ \lambda_{i} $ corresponds to the fraction of edges incident to \ac{VN}s with degree $i$.
Similarly, the \ac{CN} edge-oriented degree distribution polynomial is given by
$
\rho \left(x\right) = \sum_{i}\rho_{i}x^{i-1}
$
where $ \rho_{i} $ corresponds to the fraction of edges incident to \ac{CN}s with degree $ i $.
An unstructured irregular \ac{LDPC} code ensemble $\Cirreg$ is the set of all $q$-ary \ac{LDPC} codes with block length $ n $ and degree distributions $ \lambda\left( x\right) $ and $ \rho\left( x\right) $ and  edge labels uniformly chosen in $\Fqz$.

%% file: tex_files/lmp_description.tex
\section{\ac{SRLMP} Decoding}\label{sec:lmp_description}
This section  introduces an extension of the \ac{SMP} algorithm \cite{smp_lazaro} for transmission over a \ac{QSC}. An exchanged message between a check and a variable node is a list of symbols from $\Fq$ of size between $0$ (empty set) and  $\Gamma$, i.e., the message alphabet is $\cM_\Gamma$ which  contains all possible sets of symbols in $\Fq$ of size less \EBY{than} or equal \EBY{to} $\Gamma$ (including the empty set). The cardinality of the message alphabet is $|\cM_\Gamma| = \sum\limits_{i=0}^{\Gamma} \binom{q}{i}$. 

\begin{enumerate}[label=\roman*.]
	\item \textbf{Initialization.}
	
		Each \ac{VN} sends its channel observation $y$ to its neighboring \acp{CN}: 
		\begin{equation}
	\mes{\vn}{\cn}{0}=y.
\end{equation}
\smallskip
	\item \textbf{Check to variable update}
	
	  \EBY{Consider a \ac{CN} $\cn$ and a \ac{VN} $\vn$ connected to it. If  all of the incoming messages to $\cn$ from the other neighboring \acp{VN} are not empty, $\cn$} computes the set of all symbols that satisfy the parity check equation given the received \ac{VN} messages. Formally, \EBY{it} computes
		\begin{equation} \label{eq:CNlist}
		    \cU_{\vn,\cn}^{(\ell)}=- h_{\vn,\cn}^{-1}\sum\limits_{\vn' \in \cN (\cn) \setminus \vn} h_{\vn',\cn} \mes{\vn'}{\cn}{\ell-1}.
		\end{equation}
		 The multiplication in \eqref{eq:CNlist} is performed element-wise over $\Fq$ and the sum is over sets of symbols. The sum over two sets $\cA$ and $\cB$ is defined as \EBY{the Minkowski sum, i.e.,}
		\begin{equation}
		    \cA + \cB = \{a+b: a \in \cA, b \in \cB \}.
		\end{equation}
		If the size of $\cU_{\vn,\cn}^{(\ell)}$ is larger than $\Gamma$ or  $\cn$ receives at least one empty set from its neighboring \acp{VN}, then $\cn$ sends an empty set to $\vn$, otherwise it sends the set $\cU_{\vn,\cn}^{(\ell)}$. Formally, we write
		\begin{equation} 
		      \mes{\cn}{\vn}{\ell} = \begin{cases}
		     \cU_{\vn,\cn}^{(\ell)}  & \begin{aligned} & \text{if }\mes{\vn'}{\cn}{\ell-1} \neq \e  \, \, \forall \vn' \in \cN(\cn)\setminus \vn  \\ & \text{ and }  |\cU_{\vn,\cn}^{(\ell)}| \leq \Gamma  \end{aligned} \\ 
		    \e & \text{otherwise}.
		      \end{cases}
		\end{equation}
	\smallskip
		\item  \textbf{Variable to check update}
		
	\EBY{First, each \ac{VN}}  computes 
	\begin{equation}
		\begin{aligned} \label{eq:LL-vectorlmp}
		\aggv{\ell}&=\left[\aggc{0}{\ell},\aggc{1}{\ell},\ldots,\aggc{\alpha^{q-2}}{\ell} \right]\\ 
		&= \bm{L}\left(y\right) + \sum_{\cn' \in \neigh{\vn} \setminus \cn}  \bm{L}\left(\mes{\cn'}{\vn}{\ell}\right). 
	\end{aligned}
\end{equation}
Then, the outgoing \ac{VN} message is obtained by applying Algorithm \ref{alg:vnupdate}.
\begin{algorithm}[t]
	\begin{algorithmic}[1]
		\State Initialize the set $\cT= \emptyset$
		\State Find one symbol $a \in \Fq$ with $\aggc{a}{\ell}= \max\limits_{u \in \Fq \setminus \cT} \aggc{u}{\ell}$ \label{state:addVN}
		\State Update the set  $\cT = \cT \cup \{a\}$
		\If{$\aggc{e}{\ell} > \aggc{u}{\ell} + \Delta^{(\ell)} \, \, \forall e \in \cT $ and $\forall u \in \Fq \setminus \cT$}
		\State $  \mes{\vn}{\cn}{\ell} =\cT$
		\Else 
		\If{$|\cT|< \Gamma$}
		\State return to \ref{state:addVN}
		\Else 
		
		     $ \quad  \mes{\vn}{\cn}{\ell} =\e$
		\EndIf
		\EndIf
	\end{algorithmic}
	\caption{\ac{VN} Update Rule.}
	\label{alg:vnupdate}
\end{algorithm}
For $\Gamma=1$, Algorithm \ref{alg:vnupdate} simplifies to 
\begin{align}
 \mes{\vn}{\cn}{\ell} = \begin{cases}
 \{a\} & \begin{aligned} & \text{if } \exists a \in \Fq: \aggc{a}{\ell} > \aggc{u}{\ell} + \Delta^{(\ell)} \\&  \forall u \in \Fq \setminus \{a\}  \end{aligned} \\
 \e & \text{otherwise}
\end{cases}
\end{align}
and for $\Gamma=2$ 
\begin{align}
\mes{\vn}{\cn}{\ell} = \begin{cases}
\{a\} & \begin{aligned} & \text{if } \exists a \in \Fq: \aggc{a}{\ell} > \aggc{u}{\ell} + \Delta^{(\ell)} \\&  \forall u \in \Fq \setminus \{a\} \end{aligned} \\
\{a,e\} & \begin{aligned} &\text{if } \exists a,e \in \Fq: |\aggc{a}{\ell}  - \aggc{e}{\ell} | \leq \Delta^{(\ell)} \\& \text{and } \aggc{a}{\ell}, \aggc{e}{\ell} > \aggc{u}{\ell} +\Delta^{(\ell)} \\& \forall u \in \Fq \setminus \{a,e\} \end{aligned}\\
\e & \text{otherwise}.
\end{cases}
\end{align}
	 In \eqref{eq:LL-vectorlmp}, the \acl{llv} $\bm{L}(y)$ corresponding to the channel observation is obtained from  \eqref{eq:llvector2} using the transition probabilities of the \ac{QSC} communication channel given in \eqref{eq:QSC_trans_prob}. Further, we model each \ac{CN} to \ac{VN} message as an observation of the symbol $X$ (associated to $\vn$) at the output of an \emph{extrinsic channel}  with input alphabet $\inputalphabet= \Fq$ and output alphabet $\cZ= \cM_\Gamma$. The transition probabilities of the extrinsic channel can be estimated via \ac{DE} and are  used to compute the \aclp{llv} of the \ac{CN} messages as shown in  \eqref{eq:llvector} and \eqref{eq:llvector2}. The parameters $\Delta^{(\ell)}$ are chosen to maximize the iterative decoding threshold and can be chosen for each iteration individually.
 \smallskip
	\item \textbf{Final decision.} 
	
		Each variable node computes 
		\begin{align}
		\aggappvl{\ell}&=\left[\aggappl{0}{\ell},\aggappl{1}{\ell},\ldots,\aggappl{\alpha^{q-2}}{\ell} \right]\\ 
		&= \bm{L}\left(y\right) + \sum_{\cn' \in \neigh{\vn} } \bm{L}\left(\mes{\cn'}{\vn}{\ell}\right) .
	\end{align}
	\begin{align}\label{eq:finaldecision}
	\hat{x}^{(\ell)} = \argmax{\genfq \in \Fq}  \aggappl{\genfq}{\ell}.
\end{align}
In \eqref{eq:finaldecision}, if multiple maximizing arguments exist  we choose one of them uniformly at random.
	\end{enumerate}
Note that for $\Gamma=1$, the \ac{SRLMP} is similar to the \ac{SMP} \cite{smp_lazaro} but \ac{SRLMP} includes an additional empty set.

%% file: tex_files/lmp_DE_1.tex
\section{Density Evolution Analysis for \ac{SRLMP} with $\Gamma=1$}\label{sec:DE_lmp_1}
	This section provides a \ac{DE} analysis for \ac{SRLMP} with maximum list size $\Gamma = 1$ for non-binary irregular \ac{LDPC} code ensembles. For $\Gamma =1$, the cardinality of the  message alphabet is $|\cM_1|= q+1$. In the \ac{DE}, the probabilities of \ac{VN} to \ac{CN} and \ac{CN} to \ac{VN}  messages are tracked as iterations progress and we consider the limit as $n \to \infty$. Due to symmetry and under the all-zero codeword assumption, we can partition the message alphabet $\cM_1$ into $3$ disjoint sets $\cI_0, \cI_1, \cI_2$ such that the messages in the same set have the same probability. We have 	
	\begin{align} 
	\cI_0=&\{ \e \} \\
	\cI_1   =& \{ \{0\}\} \label{eq:setslist1} \\
	\cI_2   =& \{ \{a\}: a \in \Fqz\}.
	\end{align}	
Note that $|\cI_0|= |\cI_1| = 1 $, $|\cI_2|=q-1$. Let $\vcm{\cI_k}{\ell}$ be the probability that a \ac{VN} to \ac{CN} message belongs to the set $\cI_k$ at the $\ell$-th iteration, i.e., a \ac{VN} to \ac{CN} message takes the value $a \in \cI_k$ with probability $\vcm{\cI_k}{\ell}/| \cI_k |$. Similarly $\cvm{\cI_k}{\ell}$ is the probability that a \ac{CN} to \ac{VN} message belongs to the set $\cI_k$, where $k \in\{0,1,2\}$.

Initially, we have
	\begin{align}
\vcm{\cI_0}{0}=& 0\\
\vcm{\cI_1}{0}=& 1-\epsilon  \\
\vcm{\cI_2}{0} =& \epsilon.  
\end{align}
At the $\ell$-th iteration, we have
	\begin{align}
	\cvm{\cI_0}{\ell}=& 1- \rho\left(1- \vcm{\cI_0}{\ell-1}\right) \label{eq:s_size1I0} \\
	\begin{split}
	\cvm{\cI_1}{\ell}=&	\frac{1}{q} \Bigg[\rho\left(\vcm{\cI_1}{\ell-1}+\vcm{\cI_2}{\ell-1}\right) \\& + (q-1) \rho\left( \vcm{\cI_1}{\ell-1}-\frac{\vcm{\cI_2}{\ell-1}}{q-1}\right) \Bigg] 
	\end{split} \label{eq:s_size1I1} \\ 
		\begin{split}
	\cvm{\cI_2}{\ell}=&
\frac{q-1}{q} \Bigg[\rho\left(\vcm{\cI_1}{\ell-1}+\vcm{\cI_2}{\ell-1}\right) \\&- \rho\left(\vcm{\cI_1}{\ell-1}-\frac{\vcm{\cI_2}{\ell-1}}{q-1}\right) \Bigg]. \end{split} \label{eq:s_size1I2}
\end{align}
The extrinsic channel  has input alphabet $\inputalphabet=\Fq$, output alphabet $\cZ=\cM_1$ and transition probabilities
\begin{equation}\label{eq:transitionproblist1}
P(z|u)=\begin{cases}
\cvm{\cI_0}{\ell} & \text{if } z=\e \\ 
\cvm{\cI_1}{\ell} & \text{if } z= \{\genfq\} \\
\frac{\cvm{\cI_2}{\ell}}{q-1} & \text{if } z= \{e\} \qquad  e \in \Fq\setminus\{\genfq\}
\end{cases}
\end{equation}
Consider now the \ac{VN} to \ac{CN} messages. We define the random vector $\Fve^{(\ell)}= \left(  \Fl{0}^{(\ell)}, \ldots, \Fl{\alpha^{q-2}}^{(\ell)}, \Fe{\e}^{(\ell)} \right)$, where $\F{\a}^{(\ell)}$ denotes the \ac{RV} associated to the number of  incoming \ac{CN}  messages to a degree $d$ \ac{VN} that take value $\a \in \cM_1$ at the $\ell$-th iteration, and $\fe{\a}^{(\ell)}$  is its realization. The entries of $	\bm{L}\left(\mes{\cn'}{\vn}{\ell}\right)$ in \eqref{eq:LL-vectorlmp} are  
\begin{align}
L_u\left(\mes{\cn'}{\vn}{\ell}\right) = 	\log \left( P(\mes{\cn'}{\vn}{\ell}|u) \right)
\end{align}
	where $\mes{\cn'}{\vn}{\ell} \in \cM_1, u\in \Fq$ and $P(z|u)$ is given in \eqref{eq:transitionproblist1} $\forall z \in \cM_1$.
Hence, the elements $\aggc{u}{\ell}$ of the  extrinsic \acl{llv}  in \eqref{eq:LL-vectorlmp} are 
\begin{align}
\aggc{\genfq}{\ell} =&
\D^{(\ell)}_1 \fl{\genfq}^{(\ell)} + \D_{\tch} \delta_{\genfq y} \EBY{+ K_1}
\end{align} 
\EBY{where 
	\begin{align} 
	\begin{split}\label{eq:Klist1}
		K_1=&\log(\epsilon/(q-1))+ \fe{\e}^{(\ell)} \log(\cvm{\cI_0}{\ell})\\&+(d-1-\fe{\e}^{(\ell)}) \log(\cvm{\cI_2}{\ell} /(q-1) )
	\end{split}\\
\D_{\tch} =& \log(1-\epsilon)-\log(\epsilon/(q-1))  \label{eq:Dch}\\
\D^{(\ell)}_1=& \log(\cvm{\cI_1}{\ell})- \log(\cvm{\cI_2}{\ell}/(q-1)) \label{eq:D1}
\end{align}
and $ \delta_{ij} $ is the Kronecker delta function. Note that $K_1$ in \eqref{eq:Klist1} is independent of $\genfq$. Thus, it can be ignored when computing the extrinsic \acl{llv}.}

We obtain
\begin{align}
\begin{split}
 \vcm{\cI_0}{\ell} =& \sum\limits_{d} \lambda_d \sum_{y\in \Fq } \Pr\{Y=y\}  \sum_{\fve^{(\ell)} } \Pr\{\Fve^{(\ell)}=\fve^{(\ell)} \} \times \\ & \left[1 - \sum\limits_{a \in \Fq}\prod\limits_{\genfq \in \Fqa}\I (\aggc{a}{\ell} > \aggc{\genfq}{\ell} + \Delta^{(\ell)}) \right]
 \end{split}\\
 \begin{split}
 \vcm{\cI_1}{\ell} =& \sum\limits_{d} \lambda_d \sum_{y \in \Fq } \Pr\{Y=y\}  \sum_{\fve^{(\ell)} } \Pr\{\Fve^{(\ell)}=\fv^{(\ell)} \} \times  \\ & \prod\limits_{\genfq \in \Fqz} \I (\aggc{0}{\ell} > \aggc{\genfq}{\ell} + \Delta^{(\ell)})
 \end{split} \\ 
  \vcm{\cI_2}{\ell} =& 1-\vcm{\cI_0}{\ell}- \vcm{\cI_1}{\ell}.
    \end{align}
where   $\I(\cA)$ is an indicator function and  the inner sum is over all length $q+1$ non-negative integer vectors $\bm{f}^{(\ell)}$ whose entries  sum to $d-1$ and
\begin{align}
\Pr\{\bm{F}^{(\ell)}=\bm{f}^{(\ell)}\} =& \binom{d-1}{\fl{0}^{(\ell)},\ldots,\fe{\e}^{(\ell)}}  \prod\limits_{k=0}^{2}\left(\frac{\cvm{\cI_k}{\ell}}{| \cI_k|}\right)^{f_{\cI_k}^{(\ell)}}    \\
f_{\cI_k}^{(\ell)}=&\sum\limits_{a \in \cI_k}\f{a}^{(\ell)} \quad \forall k \in \{0,1,2\}.
\end{align}
The iterative decoding threshold $\proberror^\star$ is the maximum \EBY{ \ac{QSC}} error probability such that $ \vcm{\cI_1}{\ell} \to 1$ as $\ell \to \infty$.

%% file: tex_files/lmp_DE_2.tex
\section{Density Evolution Analysis for \ac{SRLMP} with $\Gamma=2$}\label{sec:DE_lmp_2}
This section gives a \ac{DE} analysis for \ac{SRLMP} with maximum list size $\Gamma = 2$.  For $\Gamma=2$, the cardinality of the message alphabet is $|\cM_2|= 1+ q+\binom{q}{2}$.  Due to symmetry and under the all-zero codeword assumption, we can partition the message alphabet $\cM_2$ into $5$ disjoint sets $\cI_0, \cI_1, \cI_2, \cI_3, \cI_4$ such that the messages in the same set have the same probability. We have $\cI_0, \cI_1, \cI_2$ as defined in \eqref{eq:setslist1} and	
\begin{align}
	    \cI_3=& \left\{ \{0,a\}: a \in  \Fqz \right\}\\
	    \cI_4=& \left\{ \{a,e\}: a,e \in  \Fqz \text{ and } a \neq e \right\}
	\end{align}
Note that $|\cI_0|= |\cI_1| = 1 $, $|\cI_2|=|\cI_3|=q-1$ and $|\cI_4|= \binom{q-1}{2}$.	Let $\vcm{\cI_k}{\ell}$ be the probability that a \ac{VN} to \ac{CN} message belongs to the set $\cI_k$ at the $\ell$-th iteration. Similarly $\cvm{\cI_k}{\ell}$ is the probability that a \ac{CN} to \ac{VN} message belongs to the set $\cI_k$, where $k \in\{0,1,2,3,4\}$.
	
	Initially, we have
	\begin{align}
	&\vcm{\cI_1}{0}= 1-\epsilon \\
	&\vcm{\cI_2}{0}= \epsilon\\
	&\vcm{\cI_0}{0} = \vcm{\cI_3}{0}=\vcm{\cI_4}{0}=0.  
	\end{align}
	For the \ac{CN} to \ac{VN} messages,  $\cvm{\cI_1}{\ell}$, $\cvm{\cI_2}{\ell}$, $\cvm{\cI_3}{\ell}$, $\cvm{\cI_4}{\ell}$ are given in \eqref{eq:s_size1I1}, \eqref{eq:s_size1I2},  \eqref{eq:si2} and \eqref{eq:si3}, respectively and 
	\begin{align}
	\cvm{\cI_0}{\ell}=1-\cvm{\cI_1}{\ell}-\cvm{\cI_2}{\ell}-\cvm{\cI_3}{\ell}-\cvm{\cI_4}{\ell}.
 \end{align}
	\begin{figure*}
	\begin{align}
	\begin{split}\label{eq:si2}
\cvm{\cI_3}{\ell}=&\frac{q-1}{q} \Bigg[-2\rho \left(\vcm{\cI_1}{\ell-1}+\vcm{\cI_2}{\ell-1}\right)+2\rho \left( \vcm{\cI_1}{\ell-1}+\vcm{\cI_2}{\ell-1}+\frac{\vcm{\cI_3}{\ell-1}}{q-1}+\frac{\vcm{\cI_4}{\ell-1}}{q-1}\right) 
	\\ & +(q-2) \rho\left( \vcm{\cI_1}{\ell-1}-\frac{\vcm{\cI_2}{\ell-1}}{q-1}+\frac{\vcm{\cI_3}{\ell-1}}{q-1}-\frac{2\vcm{\cI_4}{\ell-1}}{(q-1)(q-2)}\right) -(q-2)\rho\left(\vcm{\cI_1}{\ell-1}-\frac{\vcm{\cI_2}{\ell-1} }{q-1}\right)\Bigg]
\end{split}\\
\begin{split}\label{eq:si3}
\cvm{\cI_4}{\ell}=&\frac{(q-1)(q-2)}{q} \Bigg[\rho \left( \vcm{\cI_1}{\ell-1}+\vcm{\cI_2}{\ell-1}+\frac{\vcm{\cI_3}{\ell-1}}{q-1}+\frac{\vcm{\cI_4}{\ell-1}}{q-1}\right) -\rho\left(\vcm{\cI_1}{\ell-1}+\vcm{\cI_2}{\ell-1}\right) \\ & -\rho\left( \vcm{\cI_1}{\ell-1}-\frac{\vcm{\cI_2}{\ell-1}}{q-1}+\frac{\vcm{\cI_3}{\ell-1}}{q-1}-\frac{2\vcm{\cI_4}{\ell-1}}{(q-1)(q-2)}\right) +\rho\left(\vcm{\cI_1}{\ell-1}-\frac{\vcm{\cI_2}{\ell-1}}{q-1}\right)\Bigg] 
\end{split}
\end{align}
	\vspace{-2mm}
\hrule
\end{figure*}
In this case, the extrinsic channel  has input alphabet $\inputalphabet=\Fq$, output alphabet $\cZ=\cM_2$ and transition probabilities
		\begin{equation}\label{eq:transitionproblist}
	P(z|u)=\begin{cases}
	\cvm{\cI_0}{\ell} & \text{if } z= \{\genfq\} \\
	\frac{\cvm{\cI_1}{\ell}}{|\cI_1|} & \text{if } z= \{e\} \qquad  e \in \Fq\setminus\{\genfq\}\\
	\frac{\cvm{\cI_2}{\ell}}{|\cI_2|} & \text{if } z= \{\genfq,e\}\quad  e \in \Fq\setminus\{\genfq\}\\
	\frac{\cvm{\cI_3}{\ell}}{|\cI_3|} & \text{if } z= \{a,e\} \quad  a,e \in \Fq\setminus\{\genfq\}\\
	\cvm{\cI_4}{\ell} & \text{if } z=\e.
	\end{cases}
	\end{equation}
	Consider now the \ac{VN} to \ac{CN} messages. We extend the random vector $\Fv^{(\ell)}$ to 
	\[
	\Fvl^{(\ell)}= \left(  \Fl{0}^{(\ell)}, \ldots, \Fl{\alpha^{q-2}}^{(\ell)}, \Fl{0,1}^{(\ell)} , \ldots, \Fl{\alpha^{q-3},\alpha^{q-2}}^{(\ell)} ,\Fe{\e}^{(\ell)} \right),
	\]
	where $\F{\a}^{(\ell)}$ denotes the \ac{RV} associated to the number of  incoming \ac{CN}  messages to a degree $d$ \ac{VN} that take value $\a \in \cM_2$ at the $\ell$-th iteration. The entries of $	\bm{L}\left(\mes{\cn'}{\vn}{\ell}\right)$ in \eqref{eq:LL-vectorlmp} are given  by  
	\begin{align}
	L_u\left(\mes{\cn'}{\vn}{\ell}\right) = 	\log \left( P(\mes{\cn'}{\vn}{\ell}|u) \right)
	\end{align}
	where $\mes{\cn'}{\vn}{\ell} \in \cM_2, u\in \Fq$ and $P(z|u)$ is given in \eqref{eq:transitionproblist} $\forall z \in \cM_2$.
		Hence, the entries $\aggc{u}{\ell}$ of the aggregated extrinsic \acl{llv}  in \eqref{eq:LL-vectorlmp} are related to $\f{\genfq}^{(\ell)}$ and the channel observation $y$ by
\begin{align}
\aggc{\genfq}{\ell} &=
\D_1^{(\ell)} \fl{\genfq}^{(\ell)}+  \D_2^{(\ell)} \sum\limits_{a \in \Fq \setminus \{\genfq\}}\fl{\genfq,a}^{(\ell)}+ \D_{\tch} \delta_{\genfq y} \EBY{+ K_2}
\end{align} 
where $\D_{\tch}$ and $\D^{(\ell)}_1$ are given in \eqref{eq:Dch} and \eqref{eq:D1} and we have
\begin{align}
\D_2^{(\ell)}=&\log\left(\cvm{\cI_3}{\ell}/|\cI_3|\right) - \log\left(\cvm{\cI_4}{\ell}/|\cI_4|\right)\\
\begin{split} \label{eq:Klist2}
K_2=& \log(\epsilon/(q-1)) + \sum\limits_{a \in \Fq} \fl{a}^{(\ell)} \log\left(\cvm{\cI_2}{\ell}/|\cI_2|\right) \\&+ \sum\limits_{a,e \in \Fq} \fl{a,e}^{(\ell)} \log\left(\cvm{\cI_4}{\ell}/|\cI_4| \right)+ \fe{\e}^{(\ell)} \log(\cvm{\cI_0}{\ell}).
\end{split}
\end{align} 
Note that $K_2$ in \eqref{eq:Klist2} can be ignored in the \ac{VN} update rule since  it is independent of $\genfq$.

We obtain
\begin{align}
		\begin{split}
	\vcm{\cI_1}{\ell} =& \sum\limits_{d} \lambda_d \sum\limits_{y\in \Fq}\Pr\{Y=y\}\sum\limits_{\bm{f}^{(\ell)}}\Pr\{\bm{F}^{(\ell)}=\bm{f}^{(\ell)}\} \times \\& \prod\limits_{\genfq\in \Fqz}
	\I (\aggc{0}{\ell} > \aggc{\genfq}{\ell} + \Delta^{(\ell)})
		\end{split}\\
		\begin{split}
		\vcm{\cI_2}{\ell}=& \sum\limits_{d} \lambda_d\sum\limits_{a\in \Fqz} \sum\limits_{y\in \Fq}\Pr\{Y=y\}\sum\limits_{\bm{f}^{(\ell)}}\Pr\{\bm{F}^{(\ell)}=\bm{f}^{(\ell)}\}\\ &\prod\limits_{\genfq\in \Fqa}	\I (\aggc{a}{\ell} > \aggc{\genfq}{\ell} + \Delta^{(\ell)})
		\end{split} \\ 
	\begin{split}
	\vcm{\cI_3}{\ell}=& \sum\limits_{d} \lambda_d  \sum\limits_{y\in \Fq}\Pr\{Y=y\}  \sum\limits_{\bm{f}^{(\ell)}}\Pr\{\bm{F}^{(\ell)}=\bm{f}^{(\ell)}\}  \times    \\&   \sum\limits_{a\in \Fqz}   \I\left(| \aggc{0}{\ell}-\aggc{a}{\ell}| \leq \Delta ^{(\ell)} \right) \times  \\ &\prod\limits_{\genfq\in \Fq\setminus\{0,a\}}\I\left(\aggc{a}{\ell}>\aggc{\genfq}{\ell} + \Delta^{(\ell)}\right)  \I\left(\aggc{0}{\ell}> \aggc{\genfq}{\ell}+ \Delta^{(\ell)}\right)
		\end{split} \\
\begin{split}
		\vcm{\cI_4}{\ell}=& \sum\limits_{d} \lambda_d  \sum\limits_{y\in \Fq}\Pr\{Y=y\}  \sum\limits_{\bm{f}^{(\ell)}}\Pr\{\bm{F}^{(\ell)}=\bm{f}^{(\ell)}\} \times 
	 \\ & 	\sum\limits_{a,e\in \Fq \setminus \{0\},a\neq e} \mathbb{I}\left(| \aggc{a}{\ell}-\aggc{e}{\ell}| \leq \Delta^{(\ell)}\right)  \times \\ & \prod\limits_{u\in \Fq\setminus\{a,e\}}\mathbb{I}\left(\aggc{a}{\ell}> \aggc{u}{\ell}+ \Delta ^{(\ell)} \right)  \I\left(\aggc{e}{\ell}> \aggc{u}{\ell} + \Delta^{(\ell)} \right) 
\end{split} \\
	\vcm{\cI_0}{\ell}=&1-	\vcm{\cI_1}{\ell}-	\vcm{\cI_2}{\ell}-	\vcm{\cI_3}{\ell}-	\vcm{\cI_4}{\ell}
	\end{align}
	where the inner sum is over all length $|\cM_2|$ non-negative integer vectors $\bm{f}^{(\ell)}$ whose entries  sum to $d-1$
\begin{align}
 \Pr\{\bm{F}^{(\ell)}=\bm{f}^{(\ell)}\} =& \binom{d-1}{\fl{0}^{(\ell)},\ldots,\fe{\e}^{(\ell)}}  \prod\limits_{k=0}^{4}\left(\frac{\cvm{\cI_k}{\ell}}{| \cI_k|}\right)^{f_{\cI_k}^{(\ell)}}    \\
  f_{\cI_k}^{(\ell)}=&\sum\limits_{a \in \cI_k}\f{a}^{(\ell)} \quad \forall k \in \{0,\ldots,4\}.
\end{align}
	The iterative decoding threshold $\proberror^\star$ is defined as the  maximum channel error probability such that $ 	\vcm{\cI_1}{\ell} \to 1$ as $\ell \to \infty$.

%% file: tex_files/results.tex
\section{Numerical Results} \label{sec:results}
\begin{table}[t]
	\centering
	\caption{Decoding thresholds $\epsilon^\star$ of \EBY{the}  $(3,5)$ regular  LDPC code ensemble}
	\begin{tabular}{cccccccc}
		\thickhline  
		\multirow{2}{*}{$q$}&\cite{kurkoski_list}&\ac{SMP}&$\epsilon^\star$& $\epsilon^\star$ &\multirow{2}{*}{$\epsilon^\star_{\text{BP}}$}& \multirow{2}{*}{$\epsilon_{\text{Sh}}$}   \\
		& $\Gamma=1$ &\cite{smp_lazaro} &$\Gamma=1$ &$\Gamma=2$ & & \\
		\hline
		$2$  &  $0.061$ &$ 0.0611$ & $0.0975$  & $  - $        & $ 0.113$ & $ 0.1461$ \\
		
		$4$  & 0.092    &   $ 0.1226$ & $0.1283 $ & $ 0.1632 $    & $ 0.196$ & $0.2480$   \\
		
		$8$  &0.093 &   $0.1335 $ & $0.1430$  &  $ 0.1918 $   & $ 0.254$ & $ 0.3190$ \\
		
		$16$ & 0.094 &   $0.1381 $ &  $0,1627$ & $ 0.2057$     & $ 0.296$ & $  0.3708 $ \\
		
		$32$ &- &  $ 0.1403$  & $0.1906$  & $ 0.2163$     & $ 0.328$ & $0.4086$  \\
		
		$64$ & - & $ 0.1413$   & $0.2153$  & $ 0.2209$     & $ 0.352$ & $0.4369 $  \\
		\thickhline  
	\end{tabular}
	\vspace{-2mm}
	\label{tab:3,5}
\end{table}

\begin{table}[t]
	\centering
	\caption{Decoding thresholds $\epsilon^\star$ of  \EBY{the}  $(3,4)$  regular LDPC code ensemble}
	\begin{tabular}{ccccccc}
		\thickhline  
		\multirow{2}{*}{$q$}&\cite{kurkoski_list}&\ac{SMP}&$\epsilon^\star$& $\epsilon^\star$ &\multirow{2}{*}{$\epsilon^\star_{\text{BP}}$}& \multirow{2}{*}{$\epsilon_{\text{Sh}}$}   \\
		& $\Gamma=2$ &\cite{smp_lazaro} &$\Gamma=1$ &$\Gamma=2$ & & \\
		\hline
		$2$  & -     & $ 0.1069$ & $0.1439$   &$ - $    & $0.167$    &  $ 0.2145$ \\
		$4$  & 0.222 & $0.1724$  & $0.1842$   &$0.2390$ & $0.2804$   &$0.3546 $  \\
		$8$  & 0.269 & $0.1867$  & $0.2096$   &$ 0.2790$& $ 0.355$   & $0.4480 $   \\
		$16$ & 0.287 & $0.1930$  & $0.2481$   &$0.2977$ & $0.4076$   &$  0.5120$  \\
		$32$ & -     & $0.1960$  & $0.2893$   &$0.3110$ &$0.4446 $   &$ 0.5570$ \\
		$64$ & -     & $0.1974 $ & $0.3128$   &$0.3175$ &$ 0.4757$   &$ 0.5894$ \\
		\thickhline  
	\end{tabular}
	\vspace{-2mm}
	\label{tab:3,4}
\end{table} 

We investigate the asymptotic performance of \ac{SRLMP} with maximum list size $1$ and $2$ obtained by \ac{DE}. Tables \ref{tab:3,5} and \ref{tab:3,4} show the iterative decoding thresholds of \ac{SRLMP} for $(3,4)$ and $(3,5)$ regular ensembles and various values of $q$. For the sake of comparison, we provide the belief propagation thresholds $\epsilon^\star_{\text{BP}}$, the Shannon limit $\epsilon_{\text{Sh}}$ and the thresholds of the  \ac{SMP} decoder \cite{smp_lazaro}, which is similar to the \ac{SRLMP} with $\Gamma=1$ (excluding the empty set). By comparing the  thresholds for $\Gamma=1$ with the  \ac{SMP} ones, we see that significant gains are obtained if empty sets are allowed in the decoding algorithm. Increasing $\Gamma$  improves the threshold but this comes at the cost of an increasing complexity. We believe that increasing $\Gamma$ further  will significantly increase the  decoding complexity and  will not achieve significant gains compared to the case of $\Gamma=2$.
Note that the \ac{SRLMP} outperforms the decoding algorithm in \cite{kurkoski_list} for the same maximum list size.

\begin{figure}[t]
	\centering
	\footnotesize
	\includegraphics{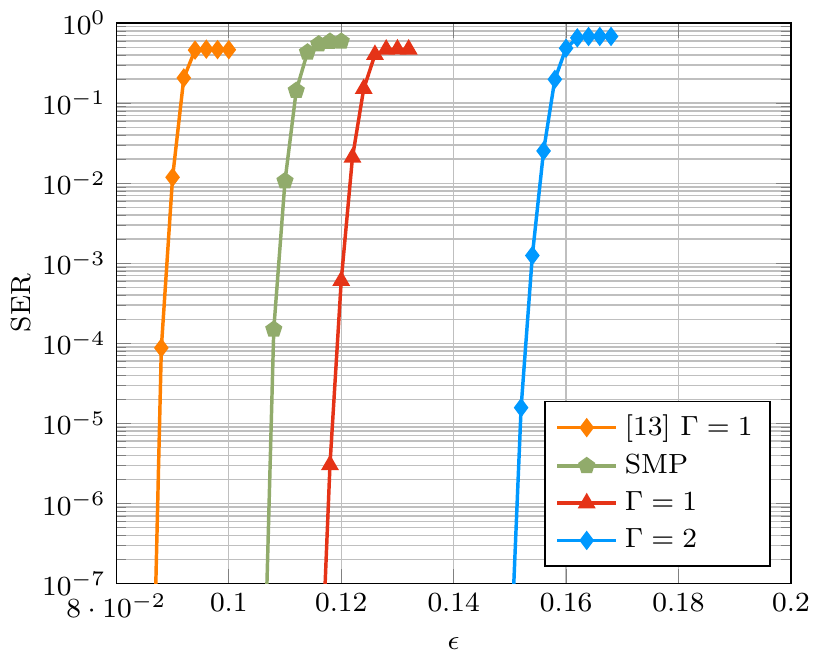}
	\caption{\ac{SER} versus channel error probability $\proberror$ for $4$-ary regular $(3,5)$ \ac{LDPC} code with $n=60000$.}
	\label{fig:results_sim_3_5q4}
	\vspace{-2mm}
\end{figure}

 Since the \ac{CN} update rule of both decoders is the same, the gain is probably due to the \ac{VN} update rule which is more complex for the case of the \ac{SRLMP} decoder. In fact, the \acp{VN} in \cite{kurkoski_list} compute the sum of binary vectors, whereas, here the incoming messages are converted to \aclp{llv} before summation.
To check the finite-length performance under \ac{SRLMP}, we consider the performance of a regular  $(3,5)$ code  where we set the maximum number of iterations $\ell_\tmax= 50$. The code has a block length $n=60000$  and its Tanner graph is obtained via the \ac{PEG} algorithm \cite{Hu_peg}. Finite-length simulation results for $\Gamma=1$ and $\Gamma=2$ are shown in Fig.\ref{fig:results_sim_3_5q4} in terms of \ac{SER} versus the \ac{QSC} error probability $\proberror$. We keep $\Delta^{(\ell)}$ constant over the iterations and  use $\Delta^{(\ell)}= 1$ for $\Gamma = 1$  and $ \Delta^{(\ell)}=1.25 $ for $\Gamma=2$.
  As a reference, we  provide the simulation results  under the \ac{SMP} decoder \cite{smp_lazaro} and under the decoding algorithm in   \cite{kurkoski_list} for $\Gamma=1$.

%% file: tex_files/conclusions.tex
\section{Conclusions}\label{sec:conclusions}
A decoding algorithm for $q$-ary \ac{LDPC} codes on the \ac{QSC} was introduced and analyzed. We presented a \ac{DE} analysis for maximum list size $1$ and $2$. The \ac{DE} yields iterative decoding thresholds and provides the reliabilities  of the extrinsic channel needed for the \ac{VN} update rule. Our algorithm outperforms competing algorithms with similar complexity.